\documentclass[useAMS,letterpaper]{mn2e}

\usepackage{epsfig}
\usepackage{times}
\usepackage{amssymb}

\newcommand{\ttms}{$t/t_{\rm MS}$}

\newcommand{\mg}{Mg\,{\sc ii}}
\newcommand{\al}{Al\,{\sc iii}}
\newcommand{\hl}{He\,{\sc i}}
\newcommand{\cl}{C\,{\sc ii}}
\newcommand{\ol}{O\,{\sc ii}}
\newcommand{\nl}{N\,{\sc ii}}
\newcommand{\kms}{km\,s$^{-1}$}
\newcommand{\vsi}{$v$\,sin\,$i$}
\newcommand{\lgg}{$\log\,g$}
\newcommand{\te}{$T_{\rm eff}$}
\newcommand{\vt}{$V_t$}

\title[The magnesium abundance in 52 B stars]
{Surface abundances of light elements for a large
sample of early B-type stars - IV. The magnesium abundance in 52
stars - a test of metallicity}
\author[L.S. Lyubimkov, S.I. Rostopchin, T.M. Rachkovskaya, D.B. Poklad and D.L. Lambert]
   {L.S.Lyubimkov$^{1}$, S.I.Rostopchin$^{1}$, T.M. Rachkovskaya$^{1}$, D.B. Poklad$^{1}$, D.L.Lambert$^{2}$ \\
$^{1}Crimean\ Astrophysical\ Observatory,\ Nauchny,\ Crimea,\
Ukraine,\ 98409$
\\ $^{2}Department\ of\ Astronomy,\ University\ of\ Texas\ at\ Austin,\ RLM\ 15.308,\
USA,\ TX\ 78712$}

\begin{document}

\date{Accepted 2004 May 15. Received 2004 May 15; in original form 2004 May 20}

\pagerange{\pageref{firstpage}--\pageref{lastpage}} \pubyear{2004}

\maketitle

\label{firstpage}

\begin{abstract}
From high-resolution spectra a non-LTE analysis of the \mg\ 4481.2
\AA\ feature is implemented for 52 early and medium local B stars
on the main sequence (MS). The influence of the neighbouring line
\al\ 4479.9 \AA\ is considered. The magnesium abundance is
determined; it is found that $\log \varepsilon({\rm
Mg})=7.67\pm0.21$ on average. It is shown that uncertainties in
the microturbulent parameter \vt\ are the main source of errors in
$\log \varepsilon({\rm Mg})$. When using 36 stars with the most
reliable \vt\ values derived from \ol\ and \nl\ lines, we obtain
the mean abundance $\log \varepsilon({\rm Mg})=7.59\pm0.15$. The
latter value is precisely confirmed for several hot B stars from
an analysis of the \mg\ 7877 \AA\ weak line. The derived abundance
$\log \varepsilon({\rm Mg})=7.59\pm0.15$ is in excellent agreement
with the solar magnesium abundance $\log \varepsilon_{\sun}({\rm
Mg})=7.55\pm0.02$, as well as with the proto-Sun abundance $\log
\varepsilon_{ps}({\rm Mg})=7.62\pm0.02$. Thus, it is confirmed
that the Sun and the B-type MS stars in our neighbourhood have the
same metallicity.
\end{abstract}


\renewcommand{\thefootnote}{}
\footnote{$^\star$ E-mail: lyub@crao.crimea.ua (LSL);
dll@astro.as.utexas.edu (DLL)}

\section{INTRODUCTION}

   In the 1990s a number of works were published, in which the carbon,
nitrogen and oxygen abundances were derived for local early B-type
stars on the main sequence (MS) (Kilian 1992; Gies \& Lambert
1992; Cunha \& Lambert 1994; Daflon et al. 1999). It was shown
that these B stars were deficient in C, N and O typically by
0.2--0.4 dex with respect to the published abundances for the Sun.

One might ask whether this deficiency reflects a real difference
in the metal abundances between the B-type MS stars and the Sun.
(It should be remembered that it is accepted in astrophysics to
denote by metals all chemical elements heavier than hydrogen and
helium). If so, then, following Nissen (1993), it should be
recognized that the metallicity of the B stars is lower by about a
factor of two than the metallicity of the Sun, i.e., the metal
mean mass fraction must be Z = 0.01 for them instead of the
usually accepted solar value Z = 0.02. An assumption arose that
the Sun formed out of material enriched by metals (see, e.g.,
Daflon et al. 1999). Moreover, a hypothesis appeared that other
F--G main sequence stars with planet systems are also more metal
rich than similar stars without planets (Gonzalez 1997, 2003).

There was a different viewpoint, namely that the initial C, N and
O abundances in the Sun and in the nearby B stars were really the
same, but the observed deficiency in the B stars results from
either the effect of gravitational-radiative separation of
elements in the stellar atmosphere or systematic errors in the
abundance analysis of the B stars.

A partial solution of the problem came unexpectedly from the
recent C, N and O abundance determinations for the Sun on the
basis of a three-dimensional time-dependent hydrodynamical model
solar atmosphere (Allende Prieto et al. 2001, 2002; Asplund 2002;
Asplund et al. 2004). It was shown that in this more realistic 3D
case the solar C, N and O abundances decrease markedly as compared
with the standard 1D case (see, e.g., Holweger 2001). So, the
discrepancy between the B stars and the Sun decreases, too.

Carbon, nitrogen and oxygen participate in the H-burning CNO-cycle
that is a main source of energy of B-type MS stars. Their abundances
in stellar interiors change significantly during the MS phase. There
are empirical data on mixing between the interiors and surface layers
of early B stars (see, e.g., Lyubimkov 1996, 1998; Maeder \& Meynet
2000). Mixing will reduce the surface C abundance and increase the
surface N abundance as material from layers exposed to the N-cycle is
brought to the surface. Mixing to deeper depths will bring products
of the ON-cycles to the surface resulting in a reduction of the
surface O abundance and further increase of the N abundance.

In fact, our analysis of the helium abundance in early B stars led to
the conclusion that the observed He enrichment during the MS phase
can be explained by rotation with velocities $v=250-400$ \kms\
(Lyubimkov et al. 2004). When adopting, for instance, $v=400$ \kms,
one may find from Heger \& Langer's (2000) computations that for a 15
Mo star the surface C, N and O abundances change by -1.2, +0.9 and
-0.4 dex, respectively. Therefore, use of C, N and O as indicators of
metallicity of early B stars is not strictly correct.

In the case of cool stars, iron is usually used as an indicator of
metallicity. On the one hand, this element does not participate in
thermonuclear reactions during early evolutionary phases, and,
therefore, its abundance corresponds to the initial metallicity of a
star. On the other hand, it gives a lot of spectral lines, which
allow its abundance to be derived with high accuracy. The metallicity
[$Fe/H$] is determined then as the difference between the Fe
abundances of the star and the Sun, i.e.

\begin{equation}
        [Fe/H] = \log \varepsilon_*({\rm Fe}) - \log \varepsilon_{\sun}({\rm Fe}),
\end{equation}

\noindent where the abundance $\log \varepsilon$ corresponds to
the standard logarithmic scale with the hydrogen abundance $\log
\varepsilon({\rm H})=12.00$. Unfortunately, in spectra of early B
stars the iron lines are weak.

One of the strongest lines in spectra of early and medium B stars
is the \mg\ 4481.2 \AA\ line. Magnesium is quite suitable as an
indicator of metallicity, because it does not change its abundance
during the MS evolutionary phase. Even its participation in the
MgAl-cycle leads to a negligible decrease of the abundance, by
about 0.02--0.03 dex (Daflon et al. 2003). Moreover, for
magnesium, unlike C, N and O, the precise solar abundance is
known. In fact, according to Holweger (2001), its abundance
derived from photospheric lines is $\log \varepsilon({\rm
Mg})=7.54\pm0.06$. This spectroscopic value is very close to the
meteoritic abundance $7.56\pm0.02$. Averaging these two
abundances, Lodders (2003) recommends the value $\log
\varepsilon_{\sun}({\rm Mg})= 7.55\pm0.02$ as the solar magnesium
abundance ('meteoritic' refers to the CI-type chondrites). We
shall use the latter value. The metallicity of stars will be
determined from formula

\begin{equation}
        [Mg/H] = \log \varepsilon_*({\rm Mg}) - \log \varepsilon_{\sun}({\rm Mg}).
\end{equation}

\noindent It should be noted that Lodders presents also the
proto-Sun magnesium abundance $\log \varepsilon_{ps}({\rm
Mg})=7.62\pm0.02$. This value differs slightly from the
above-mentioned photospheric value because heavy-element
fractionation in the Sun has altered original abundances.

At present we continue our study of large sample of early and
medium B stars. Our general goal was to provide new comprehensive
observations of the stars and to determine and analyse the
abundances of CNO-cycle elements, namely helium, carbon, nitrogen
and oxygen. A series of papers is being published. First, we
obtained high-resolution spectra of more than 100 stars at two
observatories, namely the McDonald Observatory of the University
of Texas and the Crimean Astrophysical Observatory (Ukraine);
equivalent widths of 2 hydrogen lines and 11 helium lines were
measured (Lyubimkov et al. 2000, hereinafter Paper I). Second, we
determined a number of fundamental parameters of the stars
including the effective temperature \te, surface gravity \lgg,
interstellar extinction $A_v$, distance $d$, mass $M$, radius $R$,
luminosity $L$, age $t$, and relative age \ttms, where $t_{\rm
MS}$ is the MS lifetime (Lyubimkov et al. 2002, Paper II). Third,
from non-LTE analysis of \hl\ lines we derived the helium
abundance $He/H$, microturbulent parameter \vt\ and projected
rotational velocity \vsi\ (Lyubimkov et al. 2004, Paper III).
Presently, we are analysing \cl, \nl\ and \ol\ lines; in
particular, the microturbulent parameter \vt\ is determined from
these lines, too.

We present in this paper a new determination of the Mg
abundances in B stars that is based on our spectra and our
atmospheric parameters. We use the \mg\ 4481.2 \AA\ line
and for a sample of hot stars a weaker \mg\ line at 7877 \AA.

\section{SELECTION OF STARS}

\subsection{Influence of the Al\,III 4479.9 \AA\ line}

An analysis of the \mg\ 4481.2 \AA\ line can be complicated by
presence of the neighbouring \al\ 4479.9 \AA\ line. If a star has
a rather high rotational velocity \vsi, the two lines form a
common blend. In such cases the computation of synthetic spectra
is necessary. The \al\ 4479.9 \AA\ line intensity depends strongly
on \te. In Fig.1 we show as examples the observed spectra in the
\mg\ 4481.2 \AA\ vicinity for three B stars with the different
effective temperatures \te\ and low rotational velocities \vsi\
(i.e. with sharp spectral lines). One sees that the \al\ 4479.9
\AA\ line is relatively strong at \te\ = 22500 K, but weak both at
the high temperature, \te\ = 30700 K, and at the low one, \te\ =
15800 K.

\begin{figure}
\includegraphics[width=84mm]{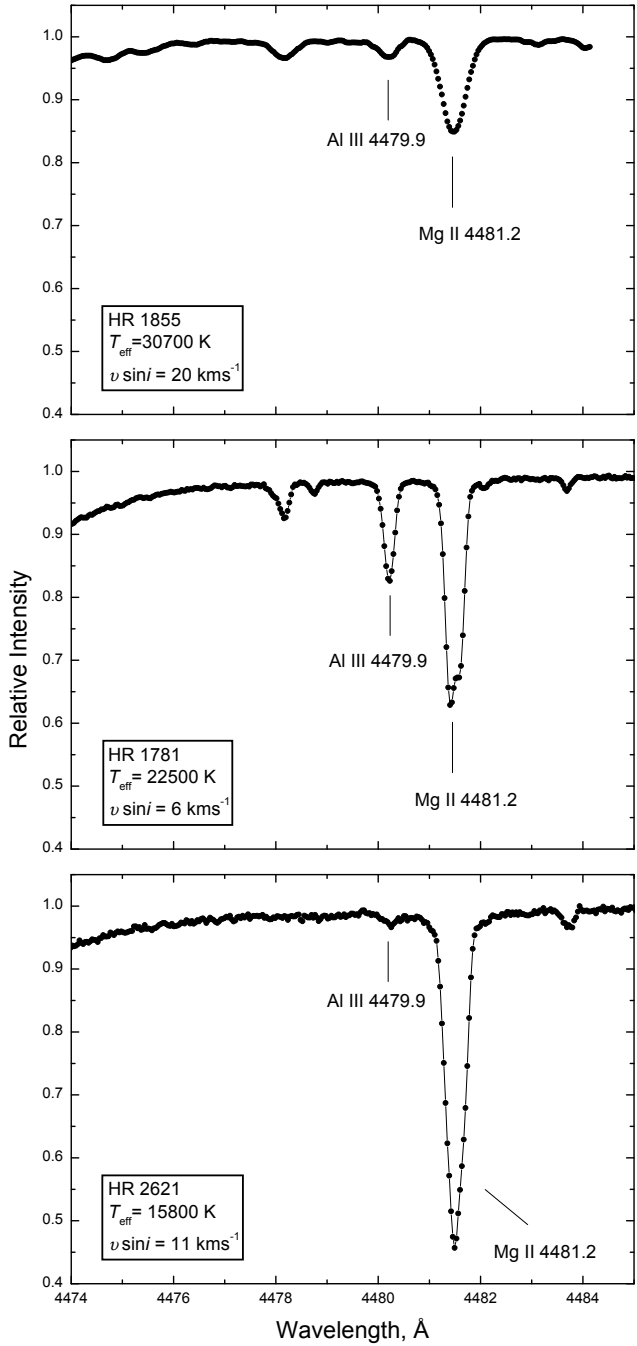}
\caption{The lines \mg\ 4481.2 \AA\ and \al\ 4479.9 \AA\ in
observed spectra of three B stars with the different effective
temperatures \te\ and low rotational velocities \vsi.}
\end{figure}

In order to study the dependence on \te\ in more detail, we
measured the equivalent width $W$ of this line for 28 B stars from
our list, which have the little rotational velocities (5 \kms\
$\leqslant$ \vsi\ $\leqslant$ 33 \kms) and the effective
temperatures \te\ $\geqslant$ 15800 K. In Fig.2a we present the
$W$ value as a function of \te. To guide the eye, we approximated
the relation by the fourth-order polynomial (solid curve in
Fig.2a). One sees that there is a maximum of the $W$ values in the
\te\ range between 21000 and 24000 K.

\begin{figure}
\includegraphics[width=84mm]{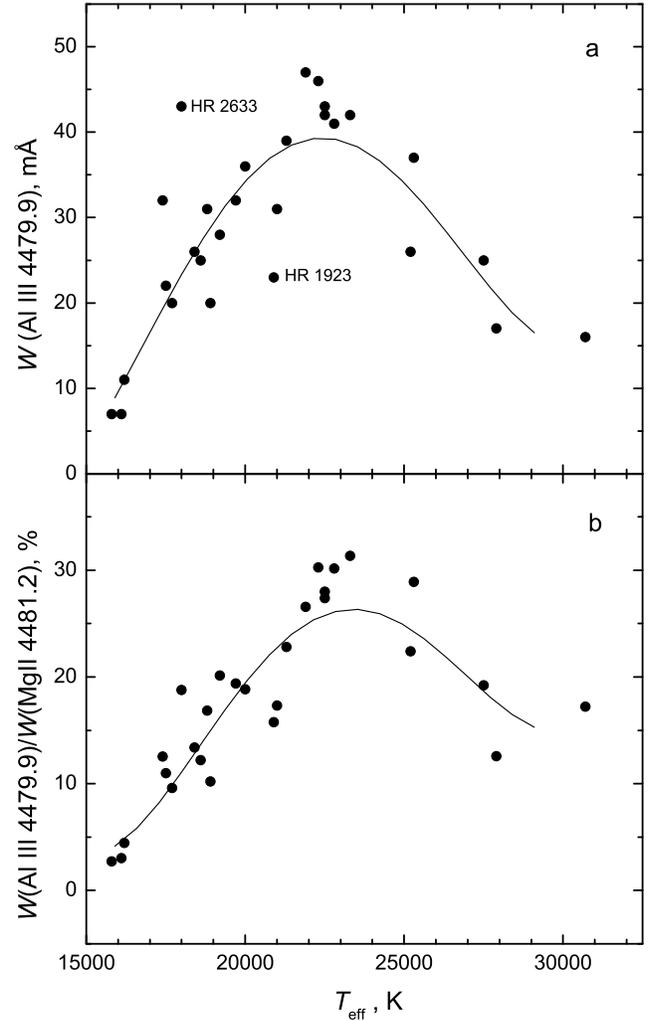}
\caption{Dependence on the effective temperature of (a) the \al\
4479.9 \AA\ line equivalent widths and (b) the ratio of the \al\
and \mg\ line equivalent widths. Solid curves correspond to
approximations by the fourth-order polynomials.}
\end{figure}

In Fig.2b we show the dependence on \te\ for the ratio of
equivalent widths of the \al\ 4479.9 \AA\ and \mg\ 4481.2 \AA\
lines. One sees that in the \te\ range from 19000 to 27000 K this
ratio is equal to about 20--30 per cent; so, if the lines form a
common blend, a contribution of the Al line must be taken into
account. We believe that the influence of the Al line can be
ignored only for B stars with temperatures \te\ $<$ 17000 K.

Two stars, namely HR\,1923 and HR\,2633, are 'outliers' in Fig.2a
(they are marked in Fig.2a). We may not explain this discrepancy
by errors in the $W$(\al\ 4479.9) measurements, because the
profiles of the Al line for these stars are sharp and symmetric
and, therefore, measured with confidence. The explanation cannot
be connected with errors in \te\ too, because they are small,
$\pm500$ and $\pm400$ K respectively. The star HR\,2633 is the
only B giant in Fig.2, it has \lgg\ = 3.38, whereas for all other
stars \lgg\ $>$ 3.8, i.e., they are dwarfs. If the \al\ 4479.9
\AA\ line is stronger for B giants than for B dwarfs with the same
temperatures \te, then the elevated $W$(\al\ 4479.9) value for
HR\,2633 becomes understandable. As far as HR\,1933 is concerned,
it is possible that this star has a somewhat lowered Al abundance.
It is interesting that in Fig.2b, unlike Fig.2a, neither of these
stars is an `outlier'.

\subsection{Selection of stars}

Although the empirical relations in Fig.2 would in principle allow
one to correct for the \al\ contribution to the \al\ - \mg\ blend
in the spectrum of a rapidly-rotating star, we chose in the
interest of controlling the uncertainties affecting the the Mg
abundance to restrict the selection of stars to those where the
\al\ and \mg\ lines are clearly resolved.

In Paper III, we determined accurate rotational velocities \vsi\
for 100 normal B stars. It was shown that the \vsi\ values of the
stars span the range from 5 to 280 \kms. In order to separate
clearly the \mg\ and \al\ lines in question, we should exclude the
stars with high \vsi. We see from our spectra that for the stars
with \vsi\ $\gtrsim$ 100 \kms\ the \mg\ line is substantially
blended by the \al\ line, especially in the above-mentioned \te\
range between 19000 and 27000 K. So, we reject
 stars with such large \vsi. Eventually, we singled out
the stars with \vsi\ $\leqslant$ 54 \kms\ when \te\ $>$ 17000 K;
one relatively cool star (\te\ = 16600 K) with \vsi\ = 70 \kms\
was added. It is important to note that in spectra of all these
stars the \al\ 4479.9 \AA\ line is distinctly separated from the
\mg\ 4481.2 \AA\ line, so equivalent widths of the latter are
measured with confidence.

A list of selected  52 stars  is presented in Table 1. We give here
for each star its HR number, as well as the effective temperature
\te\ and surface gravity \lgg\ from Paper II. Next the
microturbulent parameter \vt\ is given, which was inferred by us
for the most of the stars from \nl\ and \ol\ lines. However, for
16 cool stars these lines are too weak and, therefore, unfit for
the \vt\ determination by the standard method, when an agreement
in abundances derived from relatively strong and weak lines should
take place. We present for these stars the \vt\ values obtained in
Paper III from \hl\ lines (marked by asterisks in Table 1). The
projected rotational velocities \vsi\ from Paper III are given
next in Table 1. We remark below on the equivalent widths
$W$(4481) and magnesium abundances $\log \varepsilon({\rm Mg})$
presented in two last columns.

\begin{table}[!ht]
 \centering
 \caption{Parameters of 52 B stars, equivalent width of the \mg\ 4481 \AA\
 line and the magnesium abundance}
 \begin{tabular}{rcclrcc}
 \hline
 HR  & \te & \lgg &  \vt  & \vsi & $W$(4481) & $\log \varepsilon({\rm Mg})$\\
     &  K  &      &  \kms & \kms &  m\AA     & \\
 \hline
  38 & 18400 &  3.82 &  2.7&   11 &  194 &   7.51\\
 561 & 13400 &  3.75 &  0.0$^*$&   24 &  307 &   8.06\\
1072 & 22300 &  3.81 &  6.5&   39 &  184 &   7.57\\
1363 & 13700 &  3.67 &  1.8$^*$&   36 &  304 &   7.99\\
1595 & 22500 &  4.17 &  1.0&    6 &  157 &   7.69\\
1617 & 18300 &  3.96 &  0.5$^*$&   46 &  233 &   8.02\\
1640 & 19400 &  4.11 &  0.0&   54 &  201 &   7.82\\
1731 & 17900 &  3.98 &  2.2&   41 &  214 &   7.67\\
1753 & 15500 &  4.10 &  0.0$^*$&   27 &  274 &   7.97\\
1756 & 27900 &  4.22 &  5.1&   33 &  135 &   7.67\\
1781 & 22500 &  4.08 &  1.5&    5 &  150 &   7.60\\
1783 & 22300 &  4.00 &  1.4&   14 &  152 &   7.64\\
1810 & 20000 &  4.01 &  0.0&   24 &  191 &   7.83\\
1820 & 18600 &  4.08 &  1.0&   14 &  205 &   7.77\\
1840 & 21300 &  4.26 &  1.2&   11 &  171 &   7.67\\
1848 & 18900 &  4.23 &  0.0&   25 &  196 &   7.70\\
1855 & 30700 &  4.42 &  5.0&   20 &  93 &   7.41\\
1861 & 25300 &  4.11 &  1.7&   10 &  128 &   7.58\\
1886 & 23300 &  4.11 &  1.5&   13 &  134 &   7.49\\
1887 & 27500 &  4.13 &  4.8&   30 &  130 &   7.64\\
1923 & 20900 &  3.84 &  1.2&   17 &  146 &   7.44\\
1950 & 23100 &  4.13 &  1.8&   34 &  147 &   7.60\\
2058 & 21000 &  4.20 &  1.0&   21 &  179 &   7.75\\
2205 & 18800 &  3.74 &  2.0&    9 &  184 &   7.54\\
2222 & 25200 &  4.22 &  0.0&   17 &  116 &   7.51\\
2494 & 17500 &  4.09 &  1.6$^*$&   34 &  175 &   7.32\\
2517 & 16600 &  3.31 &  9.5&   70 &  304 &   7.43\\
2618 & 22900 &  3.39 & 12.0&   47 &  227 &   7.68\\
2621 & 15800 &  4.11 &  0.0$^*$&   11 &  258 &   7.89\\
2633 & 18000 &  3.38 &  7.0&   18 &  229 &   7.37\\
2688 & 19200 &  3.86 &  0.8&   17 &  139 &   7.24\\
2739 & 29900 &  4.10 & 10.0&   47 &  118 &   7.48\\
2756 & 16200 &  3.96 &  0.0$^*$&   26 &  248 &   7.92\\
2824 & 19400 &  3.89 &  1.2&   42 &  188 &   7.70\\
2928 & 22800 &  3.90 &  4.4&   26 &  136 &   7.31\\
3023 & 21200 &  4.02 &  1.3&   42 &  175 &   7.73\\
6588 & 17000 &  3.77 &  1.3$^*$&   45 &  205 &   7.62\\
6787 & 20000 &  3.54 &  4.2&   44 &  174 &   7.39\\
6941 & 17700 &  3.82 &  2.0$^*$&   18 &  208 &   7.64\\
6946 & 19100 &  3.76 &  0.6&   45 &  206 &   7.90\\
7426 & 16100 &  3.62 &  0.0$^*$&   29 &  230 &   7.89\\
7862 & 17100 &  4.00 &  2.0$^*$&   34 &  223 &   7.66\\
7929 & 16700 &  3.64 &  1.0$^*$&   4&  227 &   7.88\\
7996 & 15600 &  3.65 &  0.1$^*$&   35 &  251 &   8.01\\
8385 & 15000 &  3.50 &  1.2$^*$&   19 &  238 &   7.80\\
8403 & 14100 &  3.70 &  0.0$^*$&   17 &  282 &   8.03\\
8439 & 17400 &  3.31 &  5.1&   19 &  255 &   7.62\\
8549 & 19700 &  3.95 &  2.6&    8 &  165 &   7.40\\
8554 & 14000 &  4.03 &  0.0$^*$&   18 &  300 &   7.96\\
8768 & 17500 &  3.81 &  0.0&    8 &  200 &   7.70\\
8797 & 27200 &  3.98 & 10.7&   47 &  161 &   7.65\\
9005 & 21900 &  4.03 &  2.9&   15 &  177 &   7.68\\
 \hline
 \end{tabular}
 \begin{flushleft}
 $^*$ \vt\ is derived from \hl\ lines
 \end{flushleft}
\end{table}

   It should be noted that the distances of programme stars, according to
Paper II, are less than 800 pc, so the stars are really in the solar
neighbourhood.

\section{OBSERVATIONAL MATERIAL AND EQUIVALENT WIDTHS}

High-resolution spectra of programme stars have been acquired in
1996--1998 at two observatories: the McDonald Observatory (McDO)
of the University of Texas and the Crimean Astrophysical
Observatory (CrAO). At the McDO the 2.7 m telescope and coud\'{e}
echelle spectrometer (Tull et al. 1995) was used. A resolving
power was $R$ = 60000 and the typical signal-to-noise ratio was
between 100 and 300. At the CrAO we observed on the 2.6 m
telescope with coud\'{e} spectrograph. In this case we had $R$ = 30000
and a signal-to-noise ratio between 50 and 200. A more detailed
description of the observations and reductions of spectra can be
found in Paper I.

The equivalent width $W$ of the \mg\ 4481.2 \AA\ line was measured
by direct integration. The measurements were effected
independently by two of us (TMR and partially by DBP). A
comparison of these two sets of measurements showed good agreement
and an absence of a systematic difference. The averaged $W$ values
were adopted. It should be noted that a similar approach has been
used by us in Paper I in measurements of hydrogen and helium
lines. The $W$ values of the \mg\ 4481.2 \AA\ line are given in
Table 1. According to our estimates, the error in the W
determination is $\pm5$ per cent on average.

In our list there are 12 common stars, which have been observed
both at the McDO and at the CrAO. A comparison of the $W$ values
for two observatories showed good agreement: the mean difference
is $\pm4$ per cent and the maximum discrepancy is $\pm9$ per cent.
Note that the very good agreement (a scatter within $\pm5$ per
cent) has been obtained as well in Paper I for \hl\ lines.

\begin{figure}
\includegraphics[width=84mm]{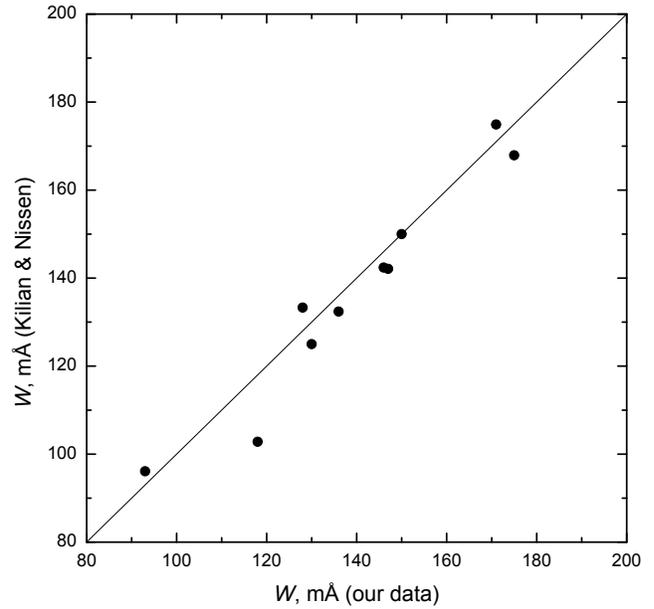}
\caption{Comparison of our measurements of the \mg\ 4481.2 \AA\
line equivalent widths with Kilian \& Nissen's (1989) data for 10
common stars.}
\end{figure}

It would be interesting to compare our $W$ values for the \mg\
4481.2 \AA\ line with other published data. Such a comparison with
measurements of Kilian \& Nissen (1989) for 10 common stars is
shown in Fig.3. One sees that for all the stars excepting one a
difference in $W$ lies within $\pm4$ per cent. The only exception
is the hot B star HR 2739, for which our equivalent width is
greater by 15 per cent than Kilian \& Nissen's value. It should be
noted in this connection that use of Kilian \& Nissen's equivalent
width for this star, according to our calculations, leads to the
magnesium abundance $\log \varepsilon({\rm Mg})=7.37$, whereas our
$W$ value gives $\log \varepsilon({\rm Mg})=7.48$ that is
substantially closer to the solar abundance $\log
\varepsilon_{\sun}({\rm Mg})=7.55$ (Lodders 2003). On the whole,
Fig.3 confirms the stated accuracy of our $W$ values.

\section{NON-LTE COMPUTATIONS OF THE Mg\,II 4481 LINE}

More than 30 years ago Mihalas (1972) first showed that a non-LTE
approach is necessary in the \mg\ 4481.2 \AA\ line analysis for
early B stars and especially O stars. According to his
calculations, the non-LTE equivalent widths $W$ of the line are
greater than the LTE ones, and the discrepancy increases with
increasing \te. Our computations confirm this conclusion;
moreover, the $W$ increment seems to be greater for B giants than
for B dwarfs. In particular, when \lgg\ = 3.5, a value
corresponding approximately to giants, the difference in $W$
between the non-LTE and LTE cases is about 5--6 per cent at \te\ =
20000--25000 K, i.e. comparable with errors in the measured W
values, whereas it exceeds 50 per cent at \te\ = 30000 K. Since
our sample includes both dwarfs and giants, we apply the non-LTE
approach to all stars.

The \mg\ 4481.2 \AA\ line was treated in our computations as a
triplet, and for its three components the same oscillator
strengths were adopted as in Daflon et al.'s (2003) work. In the
non-LTE calculations of the line we used the codes {\sc detail}
(Giddings 1981; Butler 1994) and {\sc surface} (Butler 1984). The
former code determines the atomic level populations by jointly
solving the radiative transfer and statistical equilibrium
equations. The latter one computes the emergent flux and,
therefore, the line profile.

We employed Przybilla et al.'s (2001) magnesium model atom. It
includes all Mg\,{\sc i} atomic levels with principal quantum
numbers n $\leqslant$ 9, all \mg\ levels with n $\leqslant$ 10 and
the Mg\,{\sc iii} ground state. This model atom is likely the most
complete one for magnesium today.
The computations were implemented on the basis of the LTE model
atmospheres calculated in Paper III with the code {\sc atlas9} of
Kurucz (1993).


In order to analyse the effect of the microturbulent parameter
\vt\ on the \mg\ 4481.2 \AA\ line equivalent width, we made trial
calculations of the line for some model atmospheres from Kurucz's
(1993) grid. We confirmed that the line strengthens and its
sensivity to \vt\ increases as the effective temperature \te\
decreases from 30000 to 14000 K. Therefore, the effect of errors
in \vt\ on the Mg abundance determination is largest for coolest B
stars from our list.

\section{THE MAGNESIUM ABUNDANCES}

\subsection{Errors in the derived abundances}

The derived magnesium abundances $\log \varepsilon({\rm Mg})$ are
presented in Table 1. When averaging these values for all 52
stars, we obtained the mean abundance $\log \varepsilon({\rm Mg})
= 7.67\pm0.21$. The difference with the solar value of 7.55 is
0.12 dex, i.e., substantially less than the uncertainty in the
mean value.

We estimated typical errors in the abundances, which originate
from uncertainties in the equivalent width $W$, effective
temperature \te, surface gravity \lgg, and microturbulent velocity
\vt. As mentioned above, the typical uncertainty in $W$ is $\pm5$
per cent. Contributions of three other parameters can depend on
\te, so we divided the stars into four groups as follows: \te\
$\leqslant$ 15000 K, \te\ between 15000 and 19000 K, \te\ between
19000 and 24000 K, \te\ $>$ 24000 K. In each group the mean errors
in \te\ and \lgg\ are evaluated from the individual errors (see
Paper II). As far as uncertainties in \vt\ are concerned, we
adopted the values $\pm2.0$ \kms\ and $\pm2.5$ \kms\ for stars
with \vt\ $<$ 5 \kms\ and \vt\ $\sim$ 10 \kms, respectively.

We found by this method that the full typical error in $\log
\varepsilon({\rm Mg})$ is about 0.14 dex for stars with \te\
between 19000 and 30000 K, about 0.20 dex for stars with \te\
between 15000 and 19000 K and can attain 0.30 dex for coolest
programme stars with \te\ around 14000 K. It is important to note
that the uncertainty in the microturbulent parameter \vt\ gives
the dominant contribution to the full error, namely from 40 to 65
per cent when \te\ decreases from 30000 to 19000 K and 70--80 per
cent for cooler stars. The latter conclusion will play an
important role in further discussion.

\subsection{Comparison with \boldmath $T_{\rm eff}$, $\log g$, $V_t$ and $v\,sin\,i$}

\begin{figure}
\includegraphics[width=84mm]{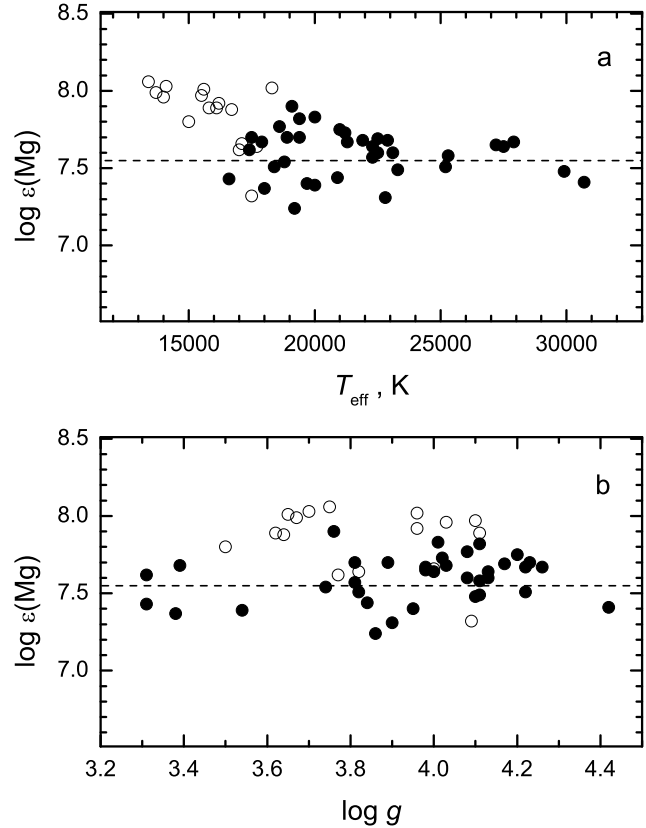}
\caption{The Mg abundance as a function of (a) the effective
temperature \te\ and (b) the surface gravity \lgg. Filled circles
correspond to stars with \vt\ from \nl\ and \ol\ lines, open
circles - stars with \vt\ from \hl\ lines. Dashed line corresponds
to the solar abundance $\log \varepsilon_{\sun}({\rm Mg}) =
7.55$.}
\end{figure}

We show in Fig.4 the derived Mg abundances as a function of the
effective temperature \te\ and surface gravity \lgg. In Fig.5 the
Mg abundances are displayed as a function of the microturbulent
parameter \vt\ and projected rotational velocity \vsi. Filled
circles correspond to 36 stars with the \vt\ values derived from
\nl\ and \ol\ lines; open circles correspond to 16 stars, where we
had to adopt the \vt\ values determined from \hl\ lines (see
above). Dashed lines correspond to the solar abundance $\log
\varepsilon_{\sun}({\rm Mg})=7.55$ (Lodders 2003).

\begin{figure}
\includegraphics[width=84mm]{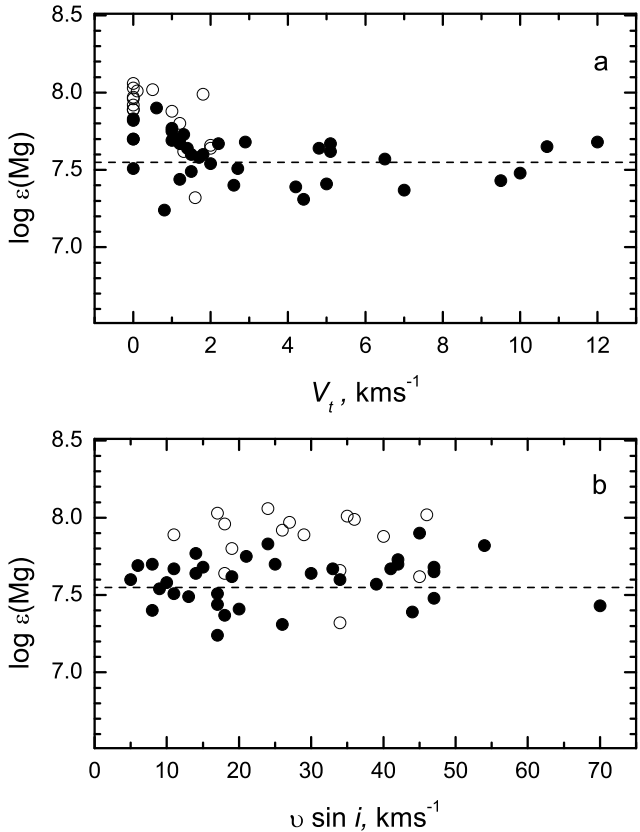}
\caption{The Mg abundance as a function of (a) the microturbulent
parameter \vt\ and (b) the projected rotational velocity \vsi.}
\end{figure}

A few points are of special interest in Fig.4 and 5, namely:

\begin{enumerate}
   \item filled circles group well around the solar Mg abundance;
   \item open circles tend to show elevated $\log \varepsilon({\rm Mg})$ values;
   \item open circles are collected in the region of lowest \te\ (Fig.4a);
   \item open circles correspond to the \vt\ values close to zero (Fig.5a).
\end{enumerate}

One may see from Fig.4a that for 11 of 16 open circles the stellar
Mg abundance exceeds the solar one by greater than 0.30 dex, i.e.,
greater than the largest error found for the coolest programme
stars (see above). These 11 stars have effective temperatures \te\
$<$ 17000 K excepting one star and the microturbulent velocities 0
$\leqslant$ \vt\ $<$ 2 \kms\ derived from \hl\ lines.

We can not explain the Mg overabundance for 11 stars by
uncertainties in the $W$ measurements. In fact, it is necessary to
lower $W$ by 15--20 per cent to eliminate the overabundance,
whereas the typical error in $W$ is 5 per cent. We considered a
systematic overestimation of $W$ due to blending of the \mg\
4481.2 \AA\ line by some other lines. According to the VALD-2
database (Kupka et al. 1999), there is only one potential line,
Fe\,{\sc ii} 4480.7 \AA, in the vicinity of \mg\ 4481.2 \AA. This
very weak line, as is the \al\ 4479.9 \AA\ line, is clearly
separated from \mg\ 4481.2 \AA\ on our spectra of cool B stars
with low \vsi.

As mentioned above, the coolest programme B stars are the most
sensitive to uncertainties in \vt; moreover, errors in \vt\ give
for them a dominating contribution, up to 80 per cent, to the
errors in $\log \varepsilon({\rm Mg})$. So, errors in \vt\ are
very likely to be a main cause of the Mg overabundance for such
stars.

We present in Table 2 the thermal velocity $V_{\rm therm}$ of the
He, N, O and Mg atoms for three temperatures T of interest. One
sees that the $V_{\rm therm}$ values for the He atoms are by about
a factor of two higher than for the N and O atoms. Even at $T$ =
15000 K, $V_{\rm therm}$ for helium is rather high, $V_{\rm
therm}$ $\approx$ 8 \kms. Therefore, when the microturbulent
parameter \vt\ is small but not zero, a broadening and
strengthening of \hl\ lines by microturbulence, unlike \nl\ and
\ol\ lines, is small compared with the effects the thermal
broadening. So, such a small \vt\ is derived from \hl\ lines with
appreciable uncertainty. One may suppose that the values \vt\
$\sim$ 0 \kms\ adopted for coolest stars are underestimated. Our
analysis showed that it is enough to increase \vt\ for such stars
from 0 to 3 or 4 \kms\ in order to arrive at agreement in $\log
\varepsilon({\rm Mg})$ with other stars.

\begin{table}
 \centering
 \caption{Thermal velocities of the He, N, O, and Mg atoms (in \kms)}
 \begin{tabular}{crrrr}
 \hline
       $T$, K &   He &    N  &   O &   Mg\\
 \hline
        15000 &  7.8 &  4.2  & 3.9 &  3.2\\
        22000 &  9.6 &  5.1  & 4.8 &  3.9\\
        30000 & 11.2 &  6.0  & 5.6 &  4.6\\
 \hline
 \end{tabular}
\end{table}

Thus, we arrive at the conclusion that the abundances $\log
\varepsilon({\rm Mg})$ represented in Fig.4 and 5 by open circles
are less reliable, because they are likely based on the
underestimated \vt\ values. When excluding these abundances, we
found the mean magnesium abundance $\log \varepsilon({\rm Mg}) =
7.59\pm0.15$ for the 36 remaining stars. This value differs from
the solar abundance only by 0.04 dex. It should be noted that it
is also very close to the proto-Sun Mg abundance $\log
\varepsilon_{ps}({\rm Mg}) = 7.62\pm0.02$ (Lodders 2003).

\subsection{Abundances from the Mg\,II 7877 \AA\ line for hot B stars}

As mentioned above, uncertainties in the microturbulent parameter
\vt\ are the main sources of errors in the Mg abundances. On our
spectra, there is a detectable \mg\ line at 7877.05 \AA, which is
rather weak in spectra of hot B stars and, therefore, insensitive
to \vt. This line is used by us as a check on the $\log
\varepsilon({\rm Mg})$ determination from the 4481 \AA\ line.

Computations of the 7877 \AA\ line with the same codes {\sc
detail} and {\sc surface} showed that for B stars with the
effective temperatures \te\ between 25000 and 30000 K its
equivalent widths $W$ vary between 15 and 30 m\AA\ and, in fact,
are almost independent of \vt. At lower \te, the 7877 \AA\ line is
stronger and sensitive to the adopted \vt. We selected from Table
1 five stars with \te\ $>$ 25000 K, namely HR\,1756, 1855, 1861,
2739, and 8797, for which we could measure the line from the McDO
spectra. The measurements were complicated by blending due to
telluric H$_2$O lines. To correct for the latter we used the
spectrum of the hot B star HR\,7446 with rapid rotation (\te\ =
26800 K, \vsi\ = 270 \kms, see Paper III). Weak  lines, like
7877 \AA, are not seen in its spectrum because of the rotation, so
the telluric spectrum can be clearly observed. When comparing this
spectrum with individual spectra of 5 stars, we allowed for the
differences in airmasses. Dividing out the telluric lines in the
vicinity of the 7877 \AA\ line, we obtained its profile and
measured the equivalent width $W$.

The observed $W$ values for 5 stars range from 21 to 29 m\AA. The
corresponding Mg abundances are very close to the values derived
from the 4481 \AA\ line. The mean Mg abundance from 7877 \AA\ for
these stars is $\log \varepsilon({\rm Mg}) = 7.59\pm0.05$, i.e.,
it coincides exactly with the above-mentioned mean abundance found
from the 4481 \AA\ line for the 36 stars with most reliable \vt.
Note that the mean Mg abundance from 4481 \AA\ for the same 5
stars is $7.56\pm0.11$. Therefore, very good agreement is found
between the Mg abundances determined from the \mg\ lines at 7877
\AA\ and 4481 \AA. Thus, we confirmed the reliability of our
results derived from the \mg\ 4481 \AA\ line.

In our further discussion, the value $\log \varepsilon({\rm Mg}) =
7.59\pm0.15$ is adopted as the final mean magnesium abundance in B
stars. Therefore, according to formula (2), the mean index of
metallicity is [$Mg/H$] = $0.04\pm0.15$. Thus, the mean
metallicity of the B-type MS stars in the solar neighborhood is
very close or equal to the metallicity of the Sun.

\subsection{Distribution of the stars on \boldmath $\log \varepsilon({\rm\bf Mg})$}

The derived magnesium abundances $\log \varepsilon({\rm Mg})$
range from 7.24 to 8.06 in the case of all 52 stars and from 7.24
to 7.90 in the case of the 36 stars with the most reliable
microturbulent parameters \vt. What are causes of the spread in
the Mg abundances? Is it possible to explain the spread only by
errors in the $\log \varepsilon({\rm Mg})$ values, or are there
some real star-to-star variations?

\begin{figure}
\includegraphics[width=84mm]{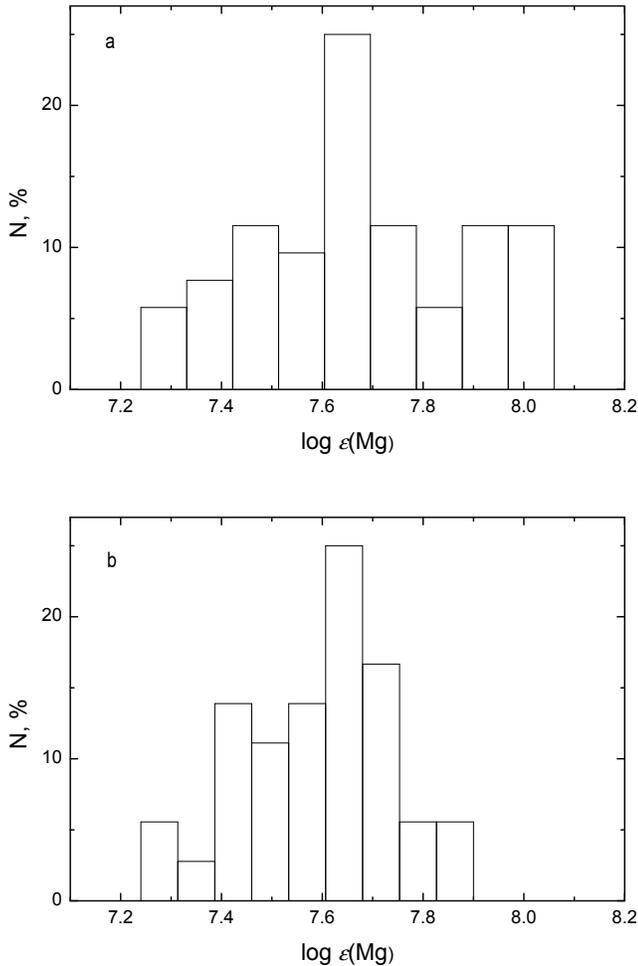}
\caption{Distribution on the Mg abundance of (a) all 52 stars and
(b) the 36 stars with most reliable \vt.}
\end{figure}

In order to answer these questions, we plotted distributions of
the abundances $\log \varepsilon({\rm Mg})$ separately for the 52
stars (Fig.6a) and the 36 stars (Fig.6b). The pronounced feature
in both histograms is the sharp maximum at $\log \varepsilon({\rm
Mg}) = 7.65$ and 7.64, respectively. It is interesting to note
that the maximum position is very close to the proto-Sun magnesium
abundance $\log \varepsilon_{ps}({\rm Mg})=7.62\pm0.02$ (Lodders
2003).

The histogram in Fig.6a shows one more feature: a second and weak
maximum between $\log \varepsilon({\rm Mg}) = 7.88$ and 8.06. This
maximum is due completely to the relatively cool B stars with
small \vt\ derived from \hl\ lines. These stars create a
substantial asymmetry in the distribution; this fact can be
considered as one more confirmation that the corresponding
abundances $\log \varepsilon({\rm Mg})$ were overestimated.

The second maximum is absent in Fig.6b, where the stars with less
reliable Vt are excluded. In Fig.6b 27 of 36 stars have effective
temperatures \te\ $>$ 19000 K; so, the typical error $\sigma$ in
$\log \varepsilon({\rm Mg})$ for them is 0.14 dex (see above). For
the remaining 9 stars with 16600 K $\leqslant$ \te\ $<$ 19000 K
the typical error is 0.20 dex. Adopting for each star the error
$\sigma$ in accordance with its \te, we found that for 34 of 36
stars a scatter of $\log \varepsilon({\rm Mg})$ around the maximum
of the distribution (i.e. $\log \varepsilon({\rm Mg}) = 7.64$) is
less than 2\,$\sigma$. Only two stars, HR\,2688 and 2928, have a
Mg underabundance that corresponds to 2.9\,$\sigma$ and
2.4\,$\sigma$. One may conclude that the scatter in Fig.6b is well
accounted for by the errors in $\log \varepsilon({\rm Mg})$.
Nevertheless, some asymmetry of the star's distribution in Fig.6b
(the left wing is more extended than the right one) may indicate
the existence of local B stars with a small Mg deficiency.

\section{DISCUSSION}

Daflon and her colleagues have analysed the \mg\ 4481 \AA\ line in
several samples of B stars with their ultimate goal being to
determine the Galactic abundance gradient of Mg from about 5 kpc to
13 kpc. Our and their analysis techniques have much in common
including the use of high-resolution high S/N spectra, use of the
codes DETAIL and SURFACE, the same Mg model atom, and the same
oscillator strengths for the components of the  4481 line. Our
methods for determining the atmospheric parameters  and the adopted
values for those parameters differ. A comparison of Mg abundances and
a discussion of the reasons for their slight difference is given in
this section.

Daflon et al. (2003) determined Mg abundances for 23 stars from
six OB associations within 1 kpc of the Sun: Galactocentric
distances ranged from 6.9 kpc to 8.2 kpc on the assumption that
the Sun is at 7.9 kpc. The mean Mg abundance for 20 stars from
the five associations with Galactocentric distances of 7.6 kpc
to 8.2 kpc is 7.37$\pm$0.14 or 0.22 dex less than our
final mean abundance.

In other papers, Daflon and colleagues (Daflon, Cunha, \& Butler 2004a,b)
obtain Mg abundances for stars inside and
outside the solar circle. Daflon \& Cunha (2004) collate the
abundances (and those of other elements) to determine and
discuss the Galactic abundance gradients. Linear fits to the
Mg abundances from 5 kpc to 13 kpc imply a solar abundance of around
7.4, i.e., a subsolar abundance but with a scatter from association
to association that brushes the solar abundance.

Systematic differences exist between our atmospheric parameters
and those used by Daflon and colleagues.  There are clear
differences in \te\ and \vt. Our \te's are about 1500--1700 K
cooler than Daflon's. What differences in the magnesium abundances
$\log \varepsilon({\rm Mg})$ can arise because of differences in
the \te\ scales? We considered as an example the star HR\,1861
with parameters \te\ = 25300 K and \lgg\ = 4.11 (Table 1) and
increased its \te\ value by 1700 K. As mentioned in Paper II, an
overestimation of effective temperatures \te\ leads automatically
to an overestimation of \lgg. (Note that traces of such
overestimation are likely seen in the \lgg\ values of Daflon et
al. (2003); unrealistically large \lgg\ are found for some stars:
\lgg\ = 4.38--4.57 whereas on the ZAMS \lgg\ $\approx$ 4.25). In
other words, if we raise \te, we should raise simultaneously \lgg.
Using the \te-\lgg\ diagram for HR 1861, we found that \lgg\ must
be increased by 0.17 dex. A redetermination of $\log
\varepsilon({\rm Mg})$ with the revised model atmosphere (\te\ =
27000 K, \lgg\ = 4.28) showed that the magnesium abundance
increases by 0.12 dex. (Note that the computed equivalent width
$W$ of the line decreases with increasing \te, but increases with
increasing \lgg, so a partial compensation takes place). We arrive
at a conclusion that the difference in the \te\ scales results in
a higher Mg abundance for Daflon et al.'s \te\ scale, but their
published abundances are lower than ours. So, the difference with
Daflon et al.'s scale cannot account for the difference in
magnesium abundance between us.

As mentioned above, uncertainties in the microturbulent parameter
\vt\ give the dominating contribution to errors in $\log
\varepsilon({\rm Mg})$. There is a significant discrepancy in the
\vt\ values between us and Daflon et al. Just this discrepancy can
be the main cause of systematic differences in $\log
\varepsilon({\rm Mg})$.

The microturbulent parameter \vt\ was derived by us for 100 B
stars from \hl\ lines in Paper III. The number of stars studied
was large, so we had an opportunity to divide the stars into three
groups according to their masses M. It was shown that in the group
of relatively low-massive stars (4.1 $\leqslant M/M_{\sun}
\leqslant$ 6.9) the \vt\ values are mostly within the 0--5 \kms\
range. The same result is obtained for stars of intermediate
masses (7.0 $\leqslant M/M_{\sun} \leqslant$ 11.2) with the
relative ages \ttms\ $<$ 0.8; however, for evolved giants with
0.80 $<$ \ttms\ $\leqslant$ 1.02 there is a large scatter in \vt\
up to \vt\ = 11 \kms. For the most massive stars of the sample
(12.4 $\leqslant M/M_{\sun} \leqslant$ 18.8) a strong correlation
between \vt\ and \ttms\ was found; for unevolved stars of the
group (\ttms\ $\leqslant$ 0.24) the velocities \vt\ vary from 4 to
7 \kms.

Very similar results were found by us recently from \nl\ and \ol\
lines. Since Daflon and colleagues  derived \vt\ from \ol\
lines, we consider at first our data inferred only from these
lines. In particular, when considering all stars of the first
group and the stars of the second group with \ttms\ $<$ 0.7, \ol\
lines give \vt\ between 0 and 5 \kms\ and a mean value \vt\ = 2.5
\kms; note that these stars have effective temperatures \te\ $<$
25500 K and surface gravities \lgg\ $\geqslant$ 3.9. For the
unevolved stars of the third group (\te\ $>$ 27000 K and \lgg\ $>$
4.0) we found from \ol\ lines, as from \hl\ lines, substantially
greater \vt\ values, 5.8 \kms\ on average. In accordance with this
finding, we selected from the lists of Daflon et al. (2003, 2004a,b)
two groups of stars, namely the stars with \te\ $<$ 25500 K and
the stars with \te\ $>$ 27000 K; in both cases \lgg\ $>$ 4.0. The
corresponding mean \vt\ values are compared with our data in Table
3.

\begin{table}
 \centering
 \caption{The mean \vt\ values (in \kms) derived from \ol\ lines for
     two groups of unevolved B stars by Daflon et al. (2003, 2004a,b) in
     comparison with our data. The number of stars used is indicated
     in bracket}
 \begin{tabular}{crrrr}
 \hline
              Source    & \te\ $<$ 25500 K  & \te\ $>$ 27000 K\\
 \hline
         Daflon et al. (2003)  &   7.0 (4)   &       8.4 (10)\\
         Daflon et al. (2004a,b)  &   4.4 (7)   &       6.5 (11)\\
         our data              &   2.5 (21)  &       5.8 (3)\\
 \hline
 \end{tabular}
\end{table}

One may see from Table 3 that there is a significant difference in
\vt\ for both groups between our and Daflon et al.'s (2003)
results, while the difference with Daflon et al. (2004a,b) is
markedly smaller. It is necessary to remember that we have used
the \vt\ values averaged on \ol\ and \nl\ lines, so our real mean
\vt\ values are somewhat smaller than in Table 3.

It is important to note once again that we determined \vt\
independently from lines of three chemical elements, namely \hl,
\nl\ and \ol. Moreover, the Vt determination from \nl\ and \ol\
lines was implemented by the standard method, but that from \hl\
lines was effected by a quite different method (Paper III).
Nevertheless, all three sets of \vt\ are in rather good agreement.
In particular, for the B stars with \te\ $<$ 25500 K we found that
(i) individual \vt\ values range, as a rule, between 0 and 5 \kms;
(ii) mean \vt\ values are 1.0, 0.5 and 2.5 \kms\ for \hl, \nl\ and
\ol, respectively. (Note that we have not found reasons to prefer
\vt(\nl) over \vt(\ol), so we used an averaged value
\vt(\nl,\ol)). Thus, our results from \hl, \nl\ and \ol\ lines
confirm that our lower \vt\ scale is preferable to the higher \vt\
scale of Daflon et al. (2003, 2004a,b). It is important to remember
as well that our mean Mg abundance 7.59 obtained with the
\vt(\nl,\ol) values from the \mg\ 4481 \AA\ line is precisely
confirmed from an analysis of the weak \mg\ 7877 \AA\ line that is
insensitive to \vt.

We implemented trial determinations of $\log \varepsilon({\rm
Mg})$ with various \vt\ for some stars from first and second
groups. The changes in $\log \varepsilon({\rm Mg})$ depend
substantially on the effective temperature \te; nevertheless, the
final conclusion is clear: the differences in the \vt\ values
explain completely the above-mentioned discrepancies
between our and Daflon et al.'s (2003) mean
magnesium abundance.

It is interesting to note that, according to Daflon (2004), the
lower \vt\ values can be a result of the cooler effective
temperatures \te. The difference between our and Daflon et al.'s
\te\ scales cannot be sufficient to cause a marked change in
thermal velocities and, hence, a change in \vt. We believe that
the difference in \vt\ discussed above may be connected with a
rough correlation between the observed equivalent widths $W$ and
excitation potentials $\chi_e$: lines with higher $\chi_e$ tend to
show lower $W$ on average. These lines are more sensitive to \te\
than weaker lines, so their calculated $W$ values change more
significantly when \te\ increases. In this case one should
increase \vt\ to eliminate a discrepancy in the derived abundances
between relatively weak and strong lines.

\section{CONCLUSIONS}

For more than ten years the C, N, and O abundances have been
considered as indicators of the metallicity of early B stars in
reference to the Sun. However, it is gradually becoming clear that
this choice is not the best one. On the one hand, there are
empirical data that mixing exists in the early B-type MS stars
between their interiors and surface layers, so the observed
abundances of the CNO-cycle elements may be affected by
evolutionary alterations. On the other hand, the solar C, N and O
abundances were continuously revised and tended to decrease during
this period. Unfortunately, it was impossible in these cases to
use the accurate meteoritic abundances, because C, N and O are
incompletely condensed in meteorites, so their meteoritic
abundances are significantly lower than the solar ones.

Magnesium, unlike C, N and O, does not have such demerits. First,
this chemical element should not alter markedly its abundance in B
stars during the MS phase. Second, its solar abundance is known
now very precisely from spectroscopic and meteoritic data.
Displaying the rather strong \mg\ 4481.2 \AA\ line in spectra of
early and medium B stars, this element is appropriate as a
reliable indicator of their metallicity.

Using the high-resolution spectra of 52 B stars we effected a
non-LTE analysis of the \mg\ 4481.2 \AA\ line and determined the
magnesium abundance. We studied the role of the neighbouring \al\
4479.9 \AA\ line and selected stars with rather low rotational
velocities \vsi\ and, therefore, with the \mg\ 4481.2 \AA\ line
well resolved from the \al\ 4479.9 \AA\ line. We found for 52
stars the mean abundance $\log \varepsilon({\rm Mg}) =
7.67\pm0.21$. The effect of uncertainties in the microturbulent
parameter \vt\ on $\log \varepsilon({\rm Mg})$ is important,
especially for coolest programme stars. For 16 such stars the \vt\
values were derived from \hl\ lines, but not from \nl\ and \ol\
lines as for other stars. Being close to zero these \vt(\hl)
values are less accurate than the \vt(\nl,\ol) values. When
excluding these 16 stars, we obtained the mean abundance $\log
\varepsilon({\rm Mg}) = 7.59\pm0.15$ for the remaining 36 stars.
This abundance is precisely confirmed from an analysis of the weak
\mg\ 7877 \AA\ line for several hot B stars. This is our
recommended Mg abundance for the B-type MS stars in the solar
neighbourhood (with $d <$ 800 pc). Comparing the latter value with
the solar magnesium abundance $\log \varepsilon_{\sun}({\rm Mg}) =
7.55\pm0.02$, one may conclude that the metallicity of the stars
is very close to the solar one. Our mean Mg abundance in B stars,
as well as  the position of the maximum in Fig.6, $\log
\varepsilon({\rm Mg}) = 7.64$, is also very close to the proto-Sun
magnesium abundance $\log \varepsilon_{ps}({\rm Mg})=7.62\pm0.02$.
We discussed the preceding determinations of the Mg abundance in B
stars by Daflon et al. (2003). Their $\log \varepsilon({\rm
Mg})$ values are somewhat lower than ours.
 We showed that this difference in $\log \varepsilon({\rm
Mg})$ is explained by differences in \vt.

Thus, our results show that the Sun is not measurably enriched in
metals as compared with the neighbouring young stars.

\section{ACKNOWLEDGEMENTS}

Two of us, LSL and SIR, are grateful to the staff of the Astronomy
Department and McDonald Observatory of the University of Texas for
hospitality during the visit in spring 2004. DLL acknowledges the
support of the Robert A. Welch Foundation of Houston, Texas.

\bsp

\label{lastpage}

\end{document}